\documentclass{ws-ijmpd}

\newcommand{\beq}{\begin{equation}}
\newcommand{\eeq}{\end{equation}}
\newcommand{\bey}{\begin{eqnarray}}
\newcommand{\eey}{\end{eqnarray}}

\newcommand{\kpc}{\, {\rm kpc} }
\newcommand{\msun}{M_\odot}

\newcommand{\grad}{{\bf \nabla}}

\newcommand{\earth}{{\oplus}}
\newcommand{\Earth}{{\oplus}}

\begin{document}

\markboth{HongSheng Zhao}
{Duality of TeVeS as both DM and DE}

%
\catchline{}{}{}{}{}
%

\title{Constraining TeVeS Gravity as Effective Dark Matter and Dark Energy}

\author{HongSheng Zhao\footnote{PPARC Advanced Fellowship}}

\address{University of St Andrews, School of Physics and Astronomy\\
KY16 9SS, Fife, UK\footnote{Member of SUPA}\\
hz4@st-and.ac.uk}



\maketitle

\begin{history}
\received{24 May 2006}
\revised{24 Sept 2006}
\comby{Managing Editor}
\end{history}

\begin{abstract}
The phenomena customly described with the standard $\Lambda$CDM model
are broadly reproduced by an extremely simple model in TeVeS, 
Bekenstein's (2004) modification of General Relativity motivated by galaxy phenomenology.  
Our model can account for the acceleration of the universe seen in SNeIa distances without 
a cosmological constant, and the accelerations seen rotation curves of nearby spiral galaxies
and gravitational lensing of high-redshift elliptical galaxies without cold dark matter. 
The model is consistent with BBN and the neutrino mass between 0.05eV to 2eV.  
The TeVeS scalar field is shown to play the effective dural roles 
of Dark Matter and Dark Energy with the amplitudes of the effects controled by 
a $\mu$-function of the scalar field, called the $\mu-$essence here.
We also discuss outliers to the theory's predictions on multi-imaged galaxy lenses
and outliers on sub-galaxy scale.  
\end{abstract}

\keywords{Dark Matter; Cosmology; Gravitation}

\section{Introduction}

As with the start of the last century, a rethinking of 
fundamental physics is forced upon us by a set of experimental surprises,
only difference with this time is that the whole universe is the laboratory.  
Einstein's General Relativity together with the ordinary matter
described by the standard model of particle physics is 
well-tested in the solar system, but
fails miserably in accounting for astronomical observations from 
just the edge of solar system to cosmological distance, e.g., 
fast-rotating galaxies like ours would have been escaped 
the shallow gravitational potentials of their luminous constituents
(stars and gas).  
Standard physics also cannot fully explain the cosmological observations of 
the cosmic acceleration seen in 
Supernovae type Ia data and 
the angular scales seen in the anisotropy spectrum 
of the Cosmic Microwave Background Radiation(CMBR) (Spergel et al. 2006).  
The remedy is usually introducing 
two exotic components to dominate the matter-energy
budget of the Universe with a split of about
$25\%:74\%$ to the universe energy budget:
Dark Matter (DM) as a colissionless and pressureless
fluid described by perhaps the SUSY physics, 
and Dark Energy (DE) as a negative pressure and nearly
homogeneous field described by unknown physics.

\subsection{Challenges to dark matter and dark energy}

In spite of the success of this concordance model 
the nature of dark matter and dark energy is
one of the greatest mysteries of modern cosmology. 
For example, it has long been noted that 
on galaxy scales dark matter and baryonic matter (stars plus gas) 
have a remarkable correlation, and respects a mysterious
acceleration scale $a_0$ (Milgrom 1983, McGaugh 2005).  
The Newtonian gravity of the baryons ${\bf g}_b$ and the dark matter gravity
${\bf g}_{DM}$ are correlated through an empirical relation
(Zhao \& Famaey 2006, Angus et al. 2006, Famaey et al. 2006) 
such that the light-to-dark ratio 
\begin{equation}\label{mueq}
 {g_b \over g_{DM} } = { g_{DM} + \alpha g_b \over a_0 }, 
\qquad a_0=1 \AA{\rm sec}^{-2}
\end{equation}
where $a_0$ is
a dividing gravity scale, and $0 \le \alpha \le 1$ is a parameter,
experimentally determined to fit rotation curves.  
\footnote{Note $\alpha$ mimics the role of the mass-to-light,
hence inherits some of its uncertainty.}
Such a tight correlation is difficult to understand in a galaxy formation
theory where dark matter and baryons interactions 
enjoy a huge degrees of freedom.  Equally peculiar is
the amplitude of dark energy density $\Lambda$, which 
is of order $10^{120}$ times smaller than its natural scale.  
It is hard to explain from fundamental physics why DE 
starts to dominate the Universe density only
at the present epoch, hence marking the present as the turning point for 
the universe from de-acceleration to acceleration.
This is related to the fact that $\Lambda \sim a_0^2$, 
where $a_0$ is a characteristic scale of DM.  Somehow DE and DM 
are tuned to shift dominance when the DM energy 
density falls below ${a_0^2 \over 8 \pi G}$.  These empirical facts 
should not be completely treated as random coincidences of 
the fundamental parameters of the universe.  

Given that the dark sector and its properties 
are only inferred indirectly from the gravitational acceleration  
of ordinary matter, one wonders if the dark sector are not 
just a sign of our lack of understanding of gravitational physics. 
Here we propose to investigate whether the role of DM and DE 
could be replaced by the scalar field in a metric theory called TeVeS.  

\section{TeVeS Framework}

TeVeS is a co-variant theory proposed by Bekenstein 
which in the weak field limit reduces to 
the phenomenogically successful but 
non-covariant MOND theory of Bekenstein \& Milgrom (1983).
The co-variant nature of TeVeS makes it ready to be analysized 
in a general setting.  

Just like Einstein's theory, Bekenstein's theory is a metric theory.
in fact, it has two metrics.  The first metric $ g_{\mu\nu}$
is minimally coupled to all the matter fields in the Universe.
We shall call the frame of this metric the ``Matter Frame'' (MF).
All geodesics are calculated in terms of this MF metric. 
For example, in a quasi-static system like a galaxy 
with a weak gravitational field, we can define a physical coordinate 
system (t,x,y,z) such that
\beq
d\tau^2 = g_{\mu\nu} dx^{\mu}dx^{\nu} 
= (1+2\Phi) dt^2 - (1-2\Phi) \left( dx^2+dy^2+dz^2\right).
\eeq
Here the potential $\Phi = \Phi_b +
\phi$ where the field $\phi$ replaces the usual role of the potential
of the Dark Matter.

Another metrics of TeVeS 
${\tilde g}_{\mu\nu}$ has its dynamics governed
by the Einstein-Hilbert action 
$S_g=\frac{1}{16\pi G}\int d^4x\sqrt{-{\tilde g}}{\tilde R},$
where ${\tilde R}$ is the scalar curvature
of ${\tilde g}_{\mu\nu}$.  
We shall call the frame of this metric the
``Einstein Frame'' (EF).  It is related to the MF metric through
${g}_{\mu\nu}=e^{-2\phi}({\tilde g}_{\mu\nu}+A_\mu A_\nu)-e^{2\phi}
A_\mu A_\nu$ (notations of tildes here are opposite of Bekenstein),
which involves the unit timelike vector field $A_{\mu}$ 
(can often be expressed as 
$(\sqrt{-g_{00}},0,0,0)$ for galaxies or FRW cosmology), and 
a scalar field $\phi$, which is governed by the action 
$S = \int d^4x  \sqrt{-{\tilde g}} {\cal L}$, where 
according to Bekenstein (2004) and Skordis et al. (2006), 
the Lagrangian density
\begin{equation}
{\cal L}= - \Lambda  + \frac{1}{16\pi G}
  \left[\mu_{Sk} {d V \over d\mu_{Sk}} - V(\mu_{Sk})\right]
\end{equation}
where $\Lambda$ is a constant of integration, equivalent to cosmological
cosntant, and $V$ is a free function of $\mu_{Sk}$, which is 
an implicit function of the scalar field $\phi$ through
\beq
\left({\tilde g}^{\mu\nu} - A^{\mu}Ae^{\nu}\right)\phi_{,\mu}\phi_{,\nu} \equiv \yen = - {d V \over d\mu_{Sk}}.
\eeq
By picking an expression and
parameters for the scalar field Lagrangian density 
${\cal L}$ or potential $V$, one picks out a given TeVeS theory.  
\footnote{These notations are directly 
related to Bekenstein's notations by $\mu_{Sk}={8 \pi \mu/k}$, 
$\yen = {y \over kl^2}$.}
Note in all above $G \equiv (1-K/2) G_\Earth$ 
is a to-be-determined bare gravitational constant 
related to the usual experimentally determined value
$G_\Earth \approx 6.67 \times 10^{-11}$ through the coupling
constant K of TeVeS (Skordis, private communication).

\section{Connecting galaxies with cosmology}

Bekenstein's original proposal was to construct the Lagrangian density 
with $ {\cal L}$ as one-to-one function of the $\mu_{Sk}$
(cf. Skordis et al 2006).
Such one-to-one construction has the drawback that the Lagrangian 
necessarily has unphysical "gaps" such that a sector is reserved for 
space-like systems (e.g. from dwarf galaxies to the solar system
in $0<\mu_{Sk}<\mu_0$) and a disconnected sector is reserved for 
time-like systems (e.g., expanding universe in $\mu_{Sk}>2\mu_0$).  
While viable mathematically, such disconnected universe would not 
permit galaxies to collapse out of the Hubble expansion.  
The particular function that Bekenstein used also result in an interpolation
function, to be computed from a non-trivial implicit function of 
the scalar field strength, which is found to overpredict 
observed rotation curve amplitudes when the gravity is of order $a_0$ 
(Famaey \& Binney 2005).  

In an effort to re-connect galaxies with the expanding universe Zhao
\& Famaey (2006) proposed to construct the Lagrangian as a one-to-one
function of the scalar field $\phi$ through $\yen$, where $\yen>0$ in
galaxies, and $\yen<0$ for cosmic expansion.  This way allows for 
a smooth transition from the edge of galaxies where $\yen \sim 0$ 
to the Hubble expansion.  Zhao \& Famaey also suggested to extrapolate
the Lagrangian for galaxies to predict cosmologies to minimize any 
fine-tuning in TeVeS.  The counterpart of the ZF model in dark matter
language would be Eq.~\ref{mueq}, which they used to fit to rotation curves,
and found that both $\alpha=1$ and $\alpha=0$ give reasonable fits, 
with some preference on the former.

Our aim here is to check whether the suggestions of Zhao \& Famaey (2006)
lead to reasonable galaxy rotation curves and cosmologies.  
To minimize fine-tuning,
we consider an extremely simple Lagrangian density governing the scalar field  
\beq
{\cal L}(\yen) = \int_0^\yen {d\yen_1 \over 8\pi G_\Earth} 
{\sqrt{|\yen|_1} \over a_0 \exp(-\phi_0)},
\qquad \Lambda=0.
\eeq
In quasi-static systems $\sqrt{|\yen|} = |\grad \phi| \exp(-\phi)$,
where the constant $\phi_0$ is the present day cosmological value of 
the scalar field $\phi$.  With this, 
the Poisson's equation reduces to 
$-\grad \left( {|\grad \exp(\phi_0-\phi)|
\over a_0 } \grad \phi \right)
= 4 \pi G_\Earth \rho = - \grad \cdot {\bf g}_b.$   
So in spherical approximations we have
\beq
{g_b \over |\grad \phi|} = {|\grad \phi| \over a(\phi)}, \qquad a(\phi) \equiv a_0 exp(\phi-\phi_0).
\eeq
Clearly the above Lagrangian or TeVeS Poisson's equation for the 
scalar field essentially recovers eq. (1) in the $\alpha=0$ case 
in the Dark Matter language
if we identify that the scalar field $\grad \phi \rightarrow g_{DM}$, hence playing the role
of gravity of the dark matter $g_{DM}$ at present day when $\phi=\phi_0$. 
Interestingly the characteristic acceleration scale $a(\phi)$ varies with redshift together
with the scalar field $\phi(t)$.  

\begin{table}[ph]
\tbl{Comparison of the pros (+) and cons (-) of LCDM vs. TeVeS in various scales}
{\begin{tabular}{@{}cccc@{}} \toprule
Data &  &   & references\\ \colrule
Pioneer Anamoly & - & + & Sanders (2006) \\
Rotation Curves HSB/LSB & -- & ++ & Zhao \& Famaey (2006) \\
Lensing by Ellipticals & ++ & +/- & Zhao, Bacon, Taylor, Horne (2006) \\
Dynamics of X-ray Clusters & ++ & +/- &  Sanders (2003), Pointecouteau \& Silk (2006) \\
Hubble expansion and CMB & ++ & +/- & Zhao et al. (2006), Skordis et al. (2005) \\ \botrule
\end{tabular} \label{ta1}}
\end{table}

A summary of how well 
TeVeS/MOND or CDM fits data on all scales is given in Table 1.
To illustrate, two sample fits to rotation curves
of a dwarf galaxy and a high surface brightness spirial galaxy 
are shown in Fig.1a,b, including the possible effects 
of imbedding the galaxies in a large neutrino core. 
We also repeat the excercise of Zhao, Bacon, Taylor \& Horne (2006), and fitting
the lens Einstein radii with Hernquist models in a $\alpha=0$ modified gravity.
We show in Fig1c,d that the CASTLES gravitational lenses (mostly high redshift ellipticals) are mostly
consistent with TeVeS predicted Einstein ring size within plausible uncertainties of the mass-to-light ratios.  
Note that the critical gravity $(c^2/D_l) (D_s/D_{ls})$ is always much stronger than $10^{-10}{\rm m/s}^2$ at 
Einstein radii of elliptical galaxy lenses, so the Einstein rings are insensitive to 
MONDian effects, hence insensitive to $a_0$.  Some of the outliers
are known to in galaxy clusters (RXJ0921 and SDS1004), where a neutrino density core of a few times
$10^{-6}\msun\kpc^{-3}$ might help to reduce the discrepency.
Given that the $\alpha=0$ model is reasonable consistent with spiral galaxy rotation curve data and 
Einstein rings of high redshift ellipticals, next we wish to study cosmology in this TeVeS model.  
The important thing to note here is that 
the cosmological constant $\Lambda$ is set to zero, so the zero point
of the Lagarangian coincide with where the scalar field is zero.

\section{Hubble expansion and late time acceleration}

TeVeS is a metric theory, the uniform expanding background can be 
described by the FRW metric.  Assume a flat cosmology with a physical time
$t$ and scale factor $a(t)$, we have
\beq
ds^2 = -dt^2 + a^2(t) 
\left[ d\chi^2 + \chi^2 \left(d\theta^2+\sin^2\theta d\phi^2\right)\right].
\eeq
The Hubble expansion can be modelled with 
\beq
\rho_\phi + \rho_b + \rho_r =  {3 H^2 \over 8 \pi G_\earth \Gamma} 
\eeq
where the first term is the scalar field effective energy density
$\rho_\phi = \yen {d {\cal L} \over d\yen} -{\cal L} =
{8 \sqrt{2} \over 3} \exp(5\phi) \left( {d \phi \over dt}\right)^3 
 \left(8 \pi G_\Earth a_0 \exp(-\phi_0) \right)^{-1}$ in the matter frame.
The correction factor 
\beq
\Gamma  \equiv  {\exp(-4\phi) \over \left( 1 + {d \phi \over d\ln a}\right)^{2} }
 \approx  \exp(4\phi_{BBN}-4\phi),
\eeq
such that the expansion rate is very close to that LCDM at the epoch of BBN
where the radiation density $\rho_r$ dominates, i.e., no corrections at BBN.
Note that  TeVeS would mimic Dark Matter and Dark Energy if we identify
\beq
\rho_\phi \Gamma \rightarrow \rho_\Lambda, \qquad
(\rho_b+\rho_r) (\Gamma-1) \rightarrow \rho_{DM}.
\eeq
where in the LCDM framework the Hubble expansion is normally modeled with 
\beq
\rho_\Lambda + \rho_{DM} + \rho_b + \rho_r = {3 H^2 \over 8 \pi G_\earth},
\qquad H={d a \over a dt}.
\eeq

Using the mirror-imaged $\alpha=0$ Lagrangian of the scalar field, 
we derive the following 2nd order ODE for the scalar field $\phi$.  
\beq
{d \over dt} \left[\mu_s \left({d \phi \over dt}\right) \right] 
= (\rho_b + \rho_r) a^3, \qquad dt = {d\ln a \over H},\qquad
\mu_s \equiv {\exp(5 \phi+\phi_0) 
a^3  \over \sqrt{2}\pi G_\Earth a_0  }  \left({d \phi \over dt}\right) 
\eeq
Note the similarity of this 1D equation with the 3D Poisson's equation.

We can integrate the above equation
to solve for $\phi$ as function of $\ln a$ or the physical synchronous time $t$.
We note at current epoch $\phi \sim \phi_0$, 
$\mu_s \sim {s \over a_0} \sim {\rho_{b} \over \rho_\phi} 
\sim \left({s^2 \over 8\pi G_\earth \rho_b}\right)^{-1/2}$, where 
$s \sim \sqrt{g_{tt}\phi_{,t}\phi_{,t}}$.  

We then aim to test if the cosmology specified by the above
non-fine-tuned Lagrangian in TeVeS could matches the behavior of a
LCDM universe.  We solve the equations numerically by iteration of the
bare constant $G$.  Assuming a value for $G$, the initial $\phi$, and
${d\phi \over d\ln a}$ are then set by the fact that $G_{eff} \approx
G_\Earth = 6.67 \times 10^{-11}$ at BBN in order to be consistent with
the number of relativistic degrees of freedom at temperature of 1MeV.
The parameters $A$ and $K$ are determined by the boundary condition at
present day such that we recover the normalisation in the MONDian
Poisson's equation in galaxies and in solar system.  Typically the
scalar field tracks the matter density, and ${\cal L}$ and $\phi$ are
slow varying functions of redshift.  We then iterate the parameter $G$
such that the sound horizon angular size at LSS ($z=1000-1100$)
matches that of LCDM.  The parameters typically converge in 20-30
iterations.  The Hubble constant and cosmic acceleration come out 
without any tuning.

In Fig.2a we show a model with a present day matter density $w_b = 0.024$.  
This is consistent with the baryon density at BBN, hence there is no 
non-baryonic matter in the present model.  This model invokes {\it neither cosmological constant 
nor dark matter}.  The resulting model has $H_0 = 77 $km/s/Mpc.
The expansion history is almost the same as LCDM; slight difference exists
in the energy density (hence the expansion rate) in the future $a>1$.

To understand whether the above explanation 
for late acceleration and dark matter 
is unique we have also run models with a more general Lagrangian.
\beq
{\cal L}(\yen) = \int_0^\yen {d\yen \over 8\pi G_\Earth} {s \over 1-\alpha s},
\qquad s \equiv {\sqrt{|\yen|} \over a_0 \exp(-\phi_0)},
\qquad \Lambda=0.
\eeq
This is so constructed that we recover eq. (1) for any value of $\alpha$
by identifying dark matter gravity with the scalar field $\grad \phi$.
This whole sequence of models are largely consistent with Dark Matter 
phenomenology on galaxy scales, with a slight preference for $\alpha=1$ 
models in galaxies.  Models with non-zero $\alpha$ also have interesting effects on the 
solar system.  For example, a model with $\alpha =0.2$ would predict 
(cf. eq. 1) a constant, non-Keperlian acceleration of 
$ a_P = a_0 \alpha^{-1} \sim 6 \times 10^{-10}{\rm m}{\rm sec}^{-2}$ in the solar system,
consistent with the Pioneer Anomaly (although an non-gravitational origin is hard to excluded). 
Such a constant gravity would cause an gravitational redshift of $10^{-13}$ (D/100AU) between
the solar system bodies of separation $D$, which could be testable with 
experiments with accurate clocks in the future (see these proceedings).  
Calculating the Hubble expansion for models with increasing $\alpha$, we are able to 
match LCDM in all cases in terms of BBN, LSS, SNeIa distances, 
and late acceleration.  
For all these models we have also varied
initial conditions and found the solutions are stable.  The acceleration 
continues into far future with $ b = a \exp(\phi) \rightarrow cst$.
The amplitude of the modification funtion $\mu$ also decreases with expansion. 
Compared to model with $\alpha=0$, however, larger $\alpha$ 
drives up the present-day Hubble constant, unless the present day 
matter density $w_b$ is also increased.  
This is effectively achieved by allowing for relatively massive neutrinos.  
E.g., for $\alpha=0.2$ would require the matter density parameter
$w_b$ twice the norminal value $0.024$, implying the need 
to include massive neutrinos of 0.8 eV.  
For $\alpha=1$ would require 2 eV neutrinos as needed for explaining galaxy cluster
data (Sanders 2003, Pointecouteau \& Silk) and the CMB (Skordis et al. 2006). 
The latter model is shown in Fig.2b.

\section{Conclusion}

In summary, we have focused on one very specific model in the
Bekenstein theory.  We have shown that it may be possible to satisfy
some of the most stringent cosmological observations without the need
to introduce/fine-tune dark matter nor dark energy.  The TeVeS scalar field $\mu$-function
(called $\mu$-essence here) can be fixed by galaxy rotation curves,
and it predicts the right amount of cosmic 
acceleration, the size of the horizon at $z=1000$, and the present Hubble constant without
fine-tuning.  The ultimate test of the model should come from simulating
the evolution of linear perturbations on this
background and CMB.  By fitting galaxy cluster data and the 3rd peak of CMB we
could break the degeneracy of models of different $\alpha$, and constrain the neutrino mass.

\section*{Acknowledgments}
I acknowledge numerous discussions with Constantinous Skordis, David Mota, Benoit Famaey.


{}

\begin{figure}
\resizebox{13cm}{!}{\includegraphics{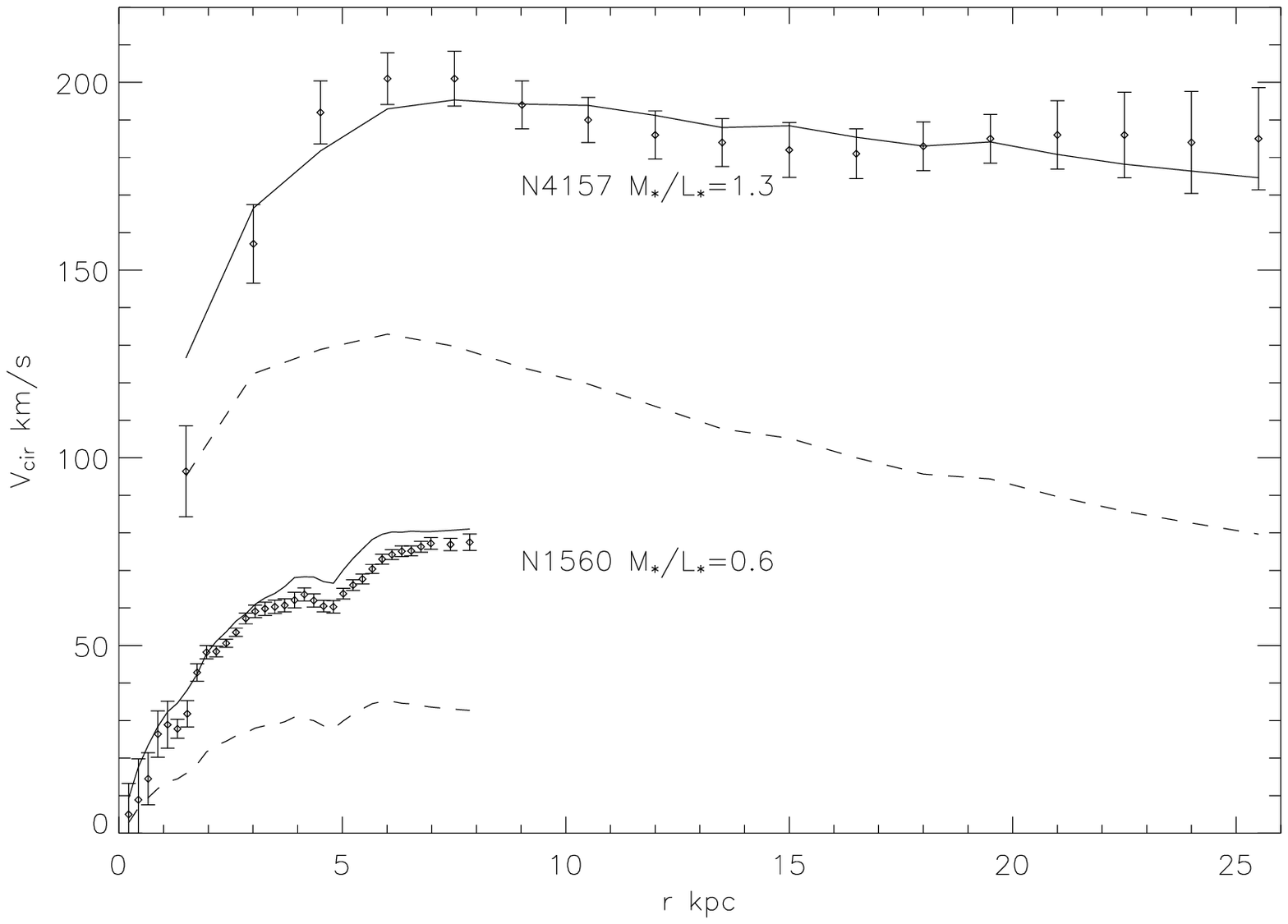},\includegraphics{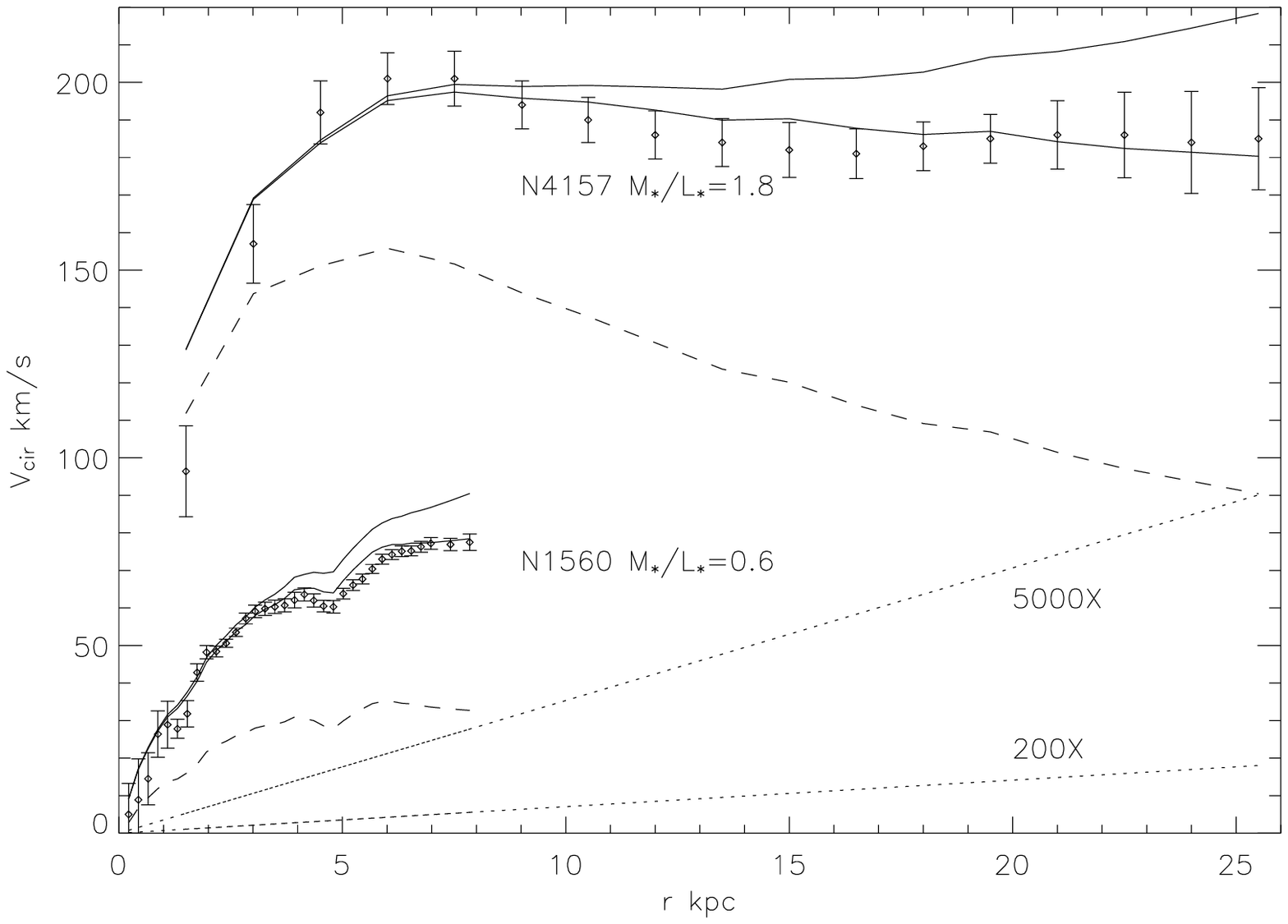}}
\caption{{\it Left panel} shows TeVeS fits to rotation curves of a gas-rich dwarf galaxy NGC1560 
and a gas-poor larger spiral galaxy NGC4157 (solid curves),
adopting $a_0=1.2\times 10^{-8}$, $\alpha=0$ $\mu$-function model without neutrinos;
the Newtonian rotation curve by baryons for the assumed stellar $(M/L)_*$ are shown as well (dashed lines).  
{\it Right panel} similar to the left, except for assuming 
the $\alpha=1$ $\mu$-function and assuming that galaxies 
are imbeded in a neutrino over-density of 
$200\times {3 H_0^2 \over 8 \pi G} \sim 2.7 \times 10^{4}\msun\kpc^{-3}$ or 
$5000\times {3 H_0^2 \over 8 \pi G} \sim 6.7 \times 10^{5}\msun\kpc^{-3}$
(the two values brackets the typical gas density of x-ray clusters on average 
and in the centres).  The Newtonian rotation curves of the constant neutrino
cores are also shown (dotted lines). }
\resizebox{13cm}{!}{\includegraphics{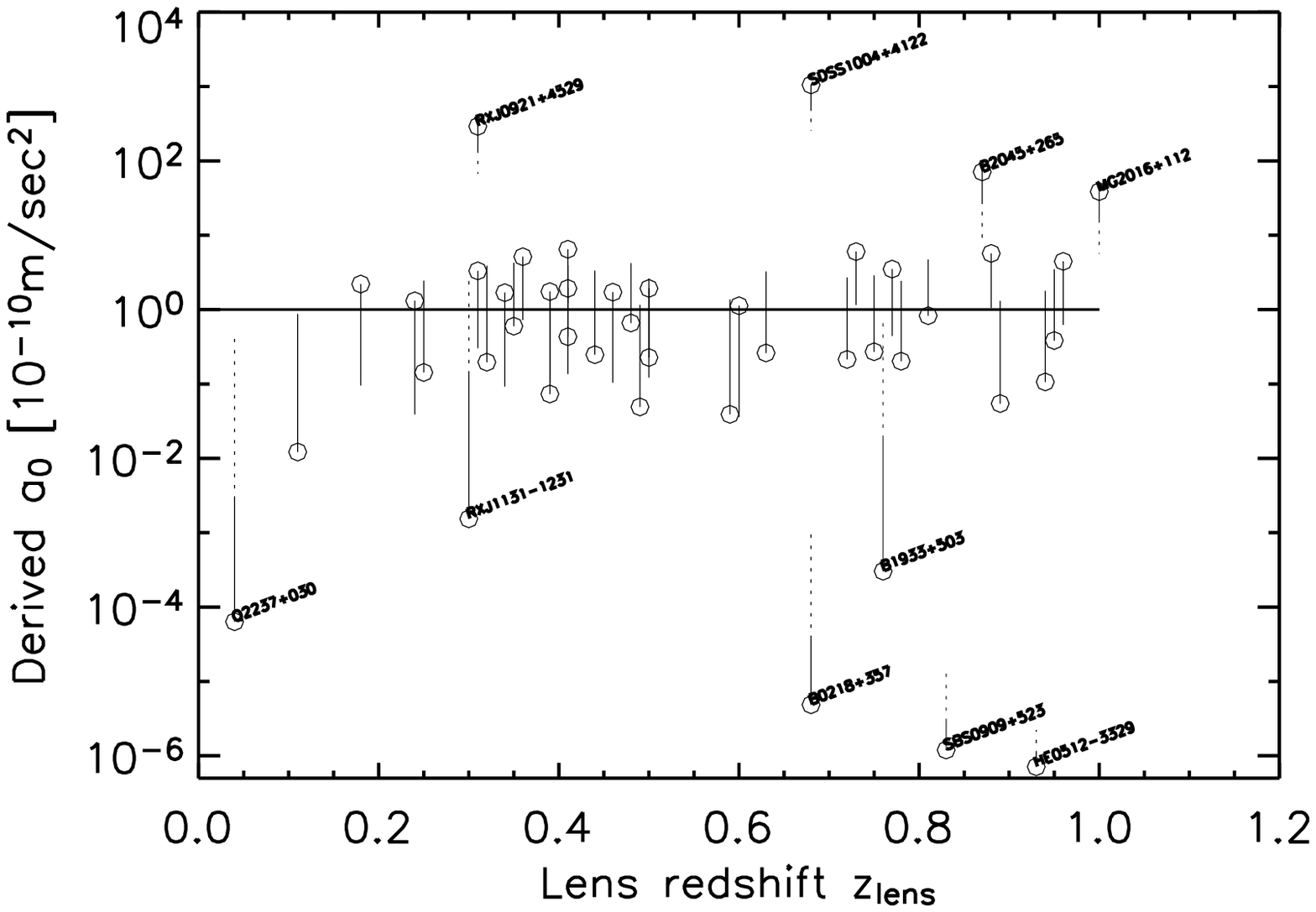},\includegraphics{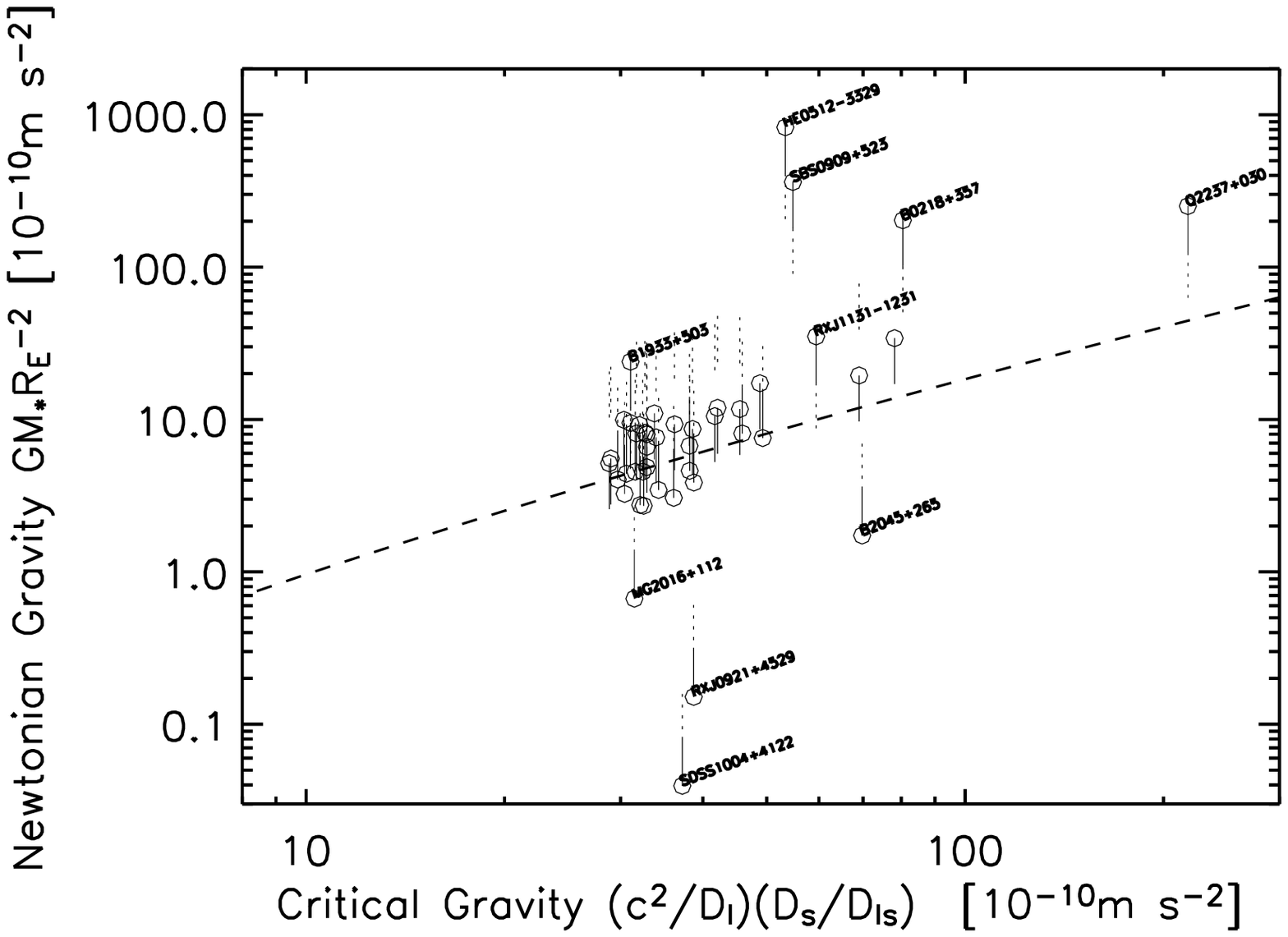}}
\caption{Shows the values for the TeVeS $a_0$ parameter derived to fit individual strong lensing Einstein radius
of 50 CASTELS multi-imaged systems, assuming $M_*/L_*=4$ (circles);
also shown are the effects of raising/lowering $M_*/L_*$ by a factor of $2$ (solid vertical lines) 
or a factor of $4$ (dotted vertical lines).  A few outliers are labeled. 
The right panel shows the Newtonian acceleration $GM_*/R_E^2$ vs. the critical 
gravity (related to the critical surface density $(c^2/4\pi G D_l) (D_s/D_{ls})$ in GR) 
is the mininal local gravity for a lens to form Einstein rings.  The dashed line is a prediction
for point lenses in TeVeS $\alpha=0$ model.}
\end{figure}

\begin{figure}
\resizebox{13cm}{!}{\includegraphics{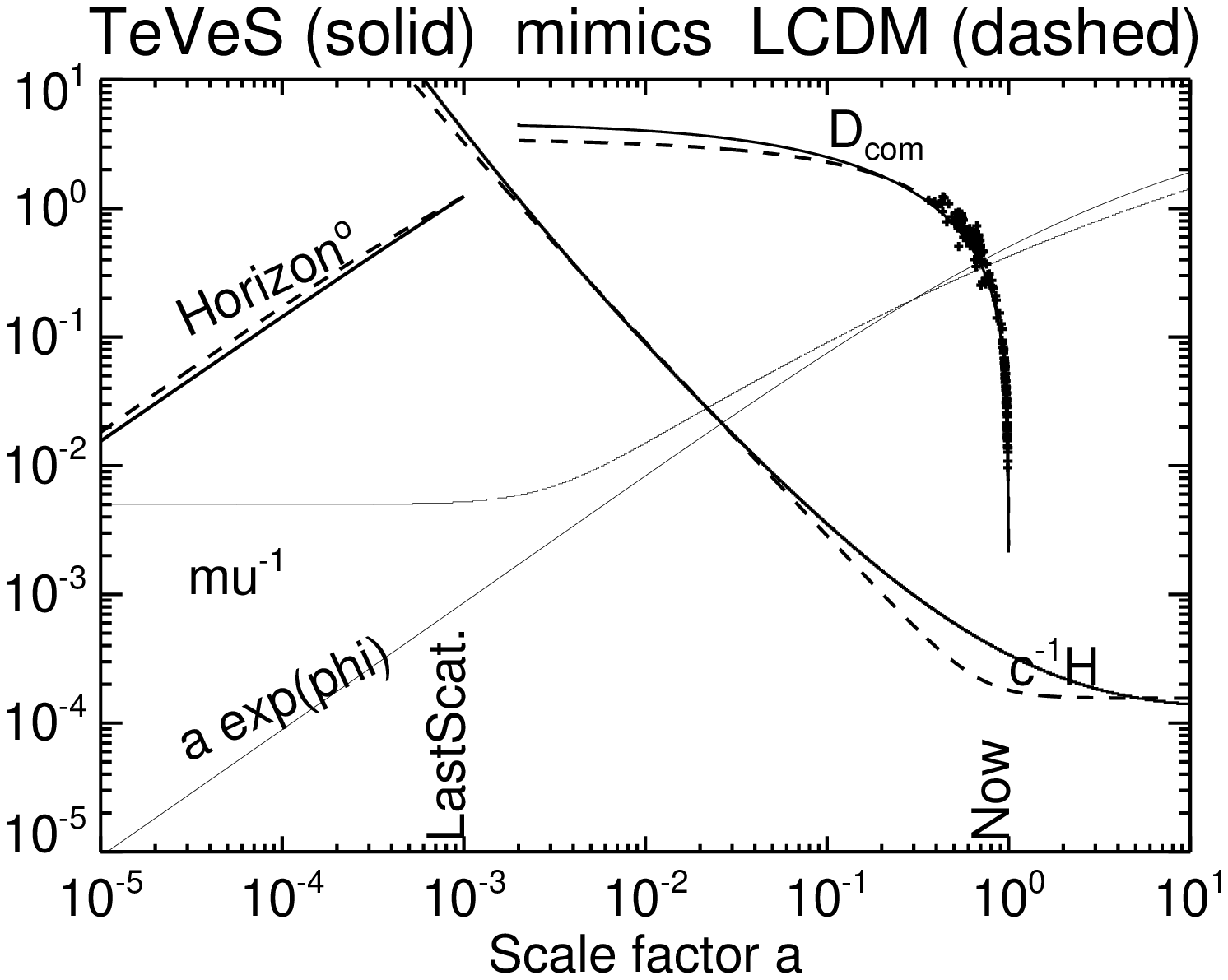},\includegraphics{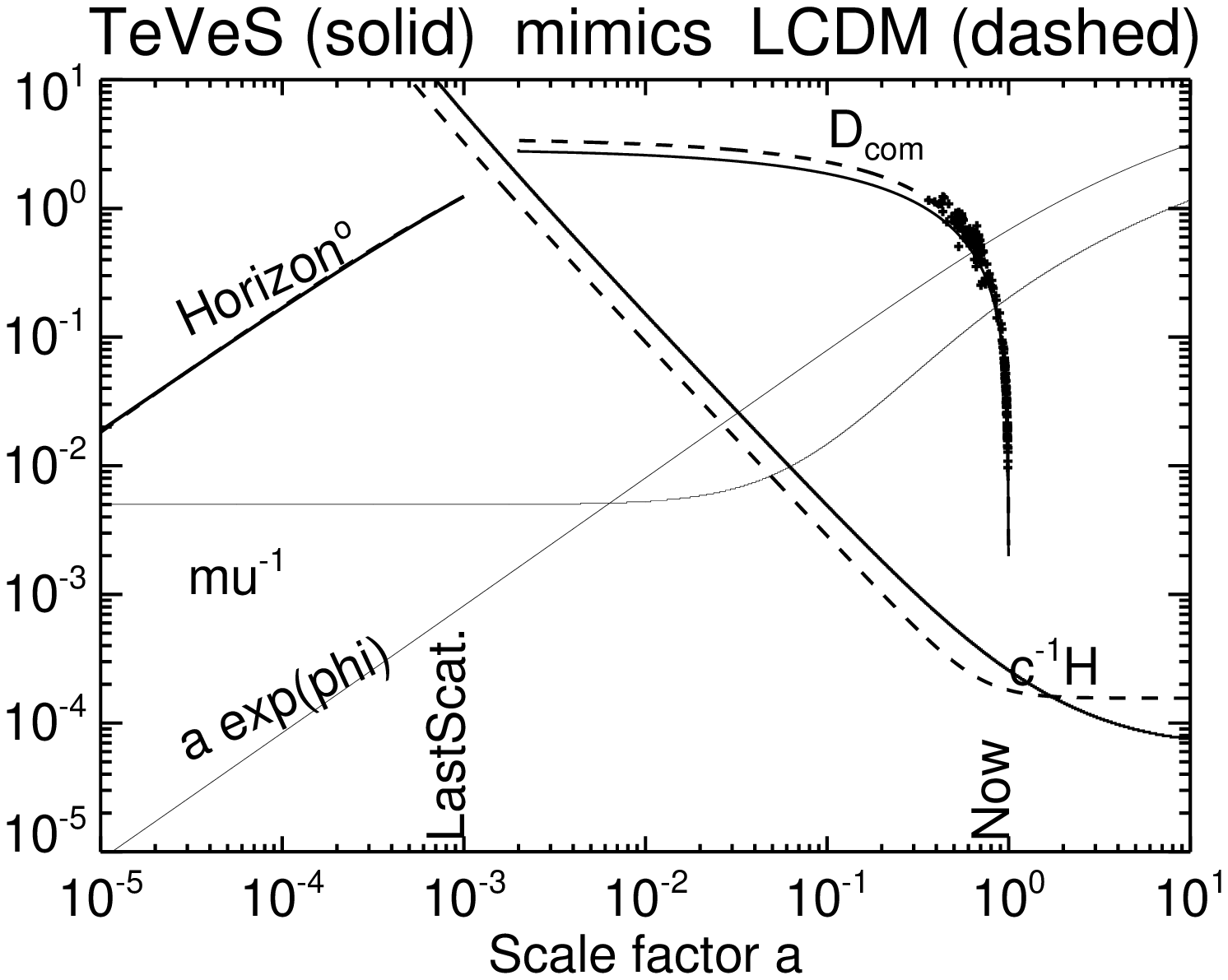}}
\caption{compares $\Lambda$CDM (dashed) with Zero-$\Lambda$-TeVeS flat cosmologies (solid) 
({\it left panel} assumes zero mass for neutrinos and a $\mu$-essence with $\alpha=0$;
{\it right panel} assumes 2eV neutrinos and $\alpha=1$ model).  Shown are
the co-moving distance $D_{com}$ 
vs. the physical scale factor $a$ in log-log diagram overplotted with 
SNIa data (small symbols) up to redshift 2.  
Likewise shows the horizon, the Hubble parameter $H$ in units of (${\rm Mpc}^{-1}c$) in two theories.    
The evolution of the scalar field $\phi$ and $\mu$  
can be inferred from (thin solid lines) $a \exp(\phi)$ and $\mu^{-1}$ with
the cutoff of $\mu^{-1}=0.005$ be adopted for numerical reasons.}
 \label{eureka}
\end{figure}

\end{document}